\documentclass[11pt]{article}
\usepackage{amsmath,amssymb}
\usepackage{graphicx,color}

\numberwithin{equation}{section}

\def\({\left(}
\def\){\right)}

\newcommand{\de}{\partial}
\newcommand{\be}{\begin{equation}}
\newcommand{\ba}{\begin{eqnarray}}
\newcommand{\ea}{\end{eqnarray}}
\newcommand{\ee}{\end{equation}}

\newcommand{\lr}{\leftrightarrow}
\newcommand{\f}{\frac}
\newcommand{\s}{\sqrt}
\newcommand{\vp}{\varphi}

\newcommand{\ti}{\tilde}
\newcommand{\ap}{\alpha}

\newcommand{\ddd}{\cdot\cdot\cdot}
\newcommand{\no}{\nonumber \\}

\newcommand{\ep}{\epsilon}

 \def\de{\partial}

 \def\lr{\leftrightarrow}
 \def\f {\frac}
 \def\ti{\tilde}
 \def\ap{\alpha}

 \def\ddd{\cdot\cdot\cdot}
 \def\no{\nonumber \\}

 \def\ep{\epsilon}

\begin{document}

\begin{titlepage}
\thispagestyle{empty}

\begin{flushright}
YITP-13-29\\
IPMU-13-0083\\
\end{flushright}

\vspace{.4cm}
\begin{center}
\noindent{\Large \textbf{Dynamics of Entanglement Entropy from
Einstein Equation}}\\
\vspace{2cm}

Masahiro Nozaki $^{a}$,
Tokiro Numasawa $^{a}$,
 Andrea Prudenziati $^{a}$,
and Tadashi Takayanagi $^{a,b}$

\vspace{1cm}
  {\it
 $^{a}$Yukawa Institute for Theoretical Physics,
Kyoto University, \\
Kitashirakawa Oiwakecho, Sakyo-ku, Kyoto 606-8502, Japan\\
\vspace{0.2cm}
 $^{b}$Kavli Institute for the Physics and Mathematics of the Universe,\\
University of Tokyo, Kashiwa, Chiba 277-8582, Japan\\
 }

\vskip 2em
\end{center}

\vspace{.5cm}
\begin{abstract}
We study the dynamics of entanglement entropy for weakly excited states in conformal field theories by using the AdS/CFT. This is aimed at a first step to find a counterpart of Einstein equation in the CFT language. In particular, we point out that the entanglement entropy satisfies differential equations which directly correspond to the Einstein equation in several setups of AdS/CFT. We also define a quantity called entanglement density in higher dimensional field theories and
study its dynamical property for weakly excited states in conformal field theories.
\end{abstract}

\end{titlepage}

\newpage


\newpage

\section{Introduction}

The AdS/CFT correspondence \cite{Maldacena,GKPW} is a remarkable fundamental relation which connects gravitational systems and quantum field theories as an equivalence. In spite of its recent successful developments, we are still far from the complete understanding of the basic mechanism of AdS/CFT correspondence (or gauge/gravity duality). The aim of the present paper is to report a modest progress in this direction. In particular we would like to
study what the Einstein equation in the gravity side corresponds to in the quantum field theory side.

Since the AdS/CFT correspondence relates gauge invariant quantities between both sides, the Einstein equation itself, which is written in term of the spacetime metric, is not directly interpreted in the dual quantum field theory. Therefore we need to find a counterpart of Einstein equation for gauge invariant quantities. We argue that
the holographic entanglement entropy (HEE) is one of the best quantities for this purpose.

The entanglement entropy in quantum field theories and more generally quantum many-body systems has been intensively studied recently (see e.g. the review articles \cite{Ereview,CCreview,CHreview,Lreview}). In AdS/CFT we can holographically calculate the entanglement entropy in a gravity dual as an area of minimal surface as conjectured in \cite{RT,HEEreview}. This holographic entanglement entropy (HEE) calculation was proved in \cite{LM} quite recently by using the bulk to boundary relation \cite{GKPW}. See \cite{CHM} for a proof when the subsystem is a round ball. Also refer to \cite{He,Ha,Fa} for strong supports towards a proof within AdS$_3/$CFT$_2$. This holographic calculation of entanglement entropy was also developed in time-dependent backgrounds \cite{HRT}, which has been applied to the quantum quenches \cite{QuenchHEE,NNT,HaMa}, the de-Sitter space \cite{MP} and energy flow \cite{NTT}. In \cite{NNT}, a falling particle in AdS was considered as the holographic dual of local quenches \cite{CaCAL} and its HEE has been computed.
In \cite{HaMa}, an analytical framework for holographic counterpart of global quantum quenches and their entanglement entropies
in CFTs \cite{CaCaG} has been discovered.
In this paper we will study how a small perturbation of HEE evolves dynamically by solving the Einstein equation in AdS spaces. A final goal will be to rewrite the perturbative
Einstein equation in terms of the HEE and we will do this explicitly in several
examples of AdS$_3/$CFT$_2$. At the same time, our results describe the behavior of entanglement entropy for weakly excited states.

Note that the complete information of HEE for arbitrary subsystems is essentially the same as that of the spacetime metric in the gravity dual (see e.g. \cite{met}). For this correspondence we do not need to know the details of the Lagrangian of matter fields coupled to the Einstein gravity. On the other hand, if we want to reproduce the spacetime metric only from the information of holographic energy stress tensor \cite{EMtensor} we need to employ the precise form of Einstein equation and thus this requires the details of matter fields.

When the size of subsystem is small, we can find a simple relation between its entanglement entropy and the total energy inside it for excited states.
This is called the first law-like relation and has been first obtained in \cite{BNTU} when excited states are static and translationally invariant.
Our analysis in this paper provides a proof of this first law relation for spherical
subsystems in the presence of time-dependent excitations.

We will also study the dynamics of entanglement density introduced in \cite{NNT} for
two dimensional field theories. We will extend this quantity in higher dimensions.
Then we consider evolutions of entanglement density imposed by the
Einstein equation.

This paper is organized as follows:
In section two we will present a general strategy of conducting our perturbative analysis of HEE and review the first law-like relation. In section three, four and five, we present our analysis of HEE in AdS$_3/$CFT$_2$, AdS$_4/$CFT$_3$ and AdS$_5/$CFT$_4$, respectively. In section six, we study the higher dimensional entanglement density and its perturbations. In section seven, we summarize our conclusions.

\section{Perturbative Calculation of Holographic Entanglement Entropy}

\subsection{General Strategy}

We consider a perturbation of the pure AdS$_{d+1}$ metric in the Fefferman-Graham (FG) gauge as follows:
\be ds^2=R^2\f{dz^2+g_{\mu\nu}(z,x)dx^\mu dx^\nu}{z^2}. \label{adsmet} \ee
The coordinate $x^\mu$ ($\mu=0,1,2,\ddd,d-1$) describes the $d$ dimensional Lorentzian space $R^{1,d-1}$ where the $d$ dimensional conformal field theory (CFT$_d$) lives in. The parameter $R$ describes the radius of the AdS space.

We consider the perturbative expansion of the form:
 \be g_{\mu\nu}=\eta_{\mu\nu}+h_{\mu\nu}, \label{pert} \ee
 assuming that $h_{\mu\nu}$ is very small. We are interested in only the linear order of $h_{\mu\nu}$. The dynamics of this perturbation is determined by the Einstein equation as usual

\be
{\cal R}_{ab}-\f{1}{2}{\cal R}g_{ab}-\f{d(d-1)}{2R^2}g_{ab}=T^{(G)}_{ab},
\label{Ein}
\ee
 where $T^{(G)}_{\mu\nu}$ is the energy stress tensor for the matter coupled to the Einstein gravity ($a,b=0,\ddd,d$ are indices of coordinates of the $d+1$ dimensional space).

The holographic entanglement entropy (HEE) in a $d+1$ dimensional AdS gravity is given in terms of the area of $d-1$ dimensional extremal surface $\gamma_A$ in a given spacetime so that the boundary of $\gamma_A$ coincides with that of the subsystem $A$ \cite{RT,HRT}. In this paper we are interested in the first order correction of HEE under a metric perturbation. This can be conveniently computed as follows. First we start with
the extremal surface $\gamma_A$ whose shape is already known in the pure AdS and calculate its area. Because the pure AdS spacetime is static, $\gamma_A$ is a minimal area surface on a canonical time slice. Next we evaluate the area of the same surface $\gamma_A$ in the perturbed metric.
The differen between these two gives precisely the first order correction of HEE
$\Delta S_A$. Thus we do not need to know how the shape of the extremal surface is modified under the metric perturbation. This simply stems from the fact that $\gamma_A$ satisfies the extremal surface condition in the pure AdS space.

Therefore we can calculate the shift of HEE $\Delta S_A$ due to the metric perturbation (\ref{pert}) from the formula  \be \Delta S_A=\f{1}{8G_N}\int
(d\zeta)^{d-1} \s{G^{(0)}}G^{(1)}_{\ap\beta}G^{(0)\ap\beta}, \label{dsa} \ee
where $\zeta$ is a coordinate of the $d-1$ dimensional extremal surface $\gamma_A$ as
employed in \cite{NNT}. $G^{(0)}$ and $G^{(1)}$ represents the induced metric on $\gamma_A$ with respect to the pure AdS and its first order perturbation. In this paper we consider examples where the subsystem $A$ is given by a round ball for which we know the analytical expression of the surface $\gamma_A$ in the pure AdS.

\subsection{First Law-like Relation}

A useful property which is enjoyed by the perturbed HEE is the first law-like relation
\cite{BNTU}. It takes the following form:
\be
T_{eff}\cdot \Delta S_A=\Delta E_A. \label{first}
\ee
When we choose $A$ to be a round ball with radius $l$, the entangling temperature $T_{eff}$ takes the universal value
\be\label{teff}
T_{eff}=\f{d+1}{2\pi l}.
\ee
Also $\Delta E_A$ denotes the total energy in the region $A$ and is written as
\be
\Delta E_A=\int_A T_{tt},
\ee
where $T_{tt}$ is the energy density in CFT$_d$. If we perform the expansion of $h_{\mu\nu}$ defined in (\ref{pert}) as
\be
h_{\mu\nu}=z^d H_{\mu\nu}+\ddd, \label{expa}
\ee
in the near AdS boundary limit $z\to 0$, the holography energy stress tensor
can be obtained as follows \cite{EMtensor}
\be
T_{\mu\nu}=\f{dR^{d-1}}{16\pi G_N}H_{\mu\nu}. \label{EMtensor}
\ee

This relation (\ref{first}) was originally confirmed in static, isotropic and translationally invariant perturbations of the metric. Therefore it is intriguing to see if it holds without this assumptions. This is another motivation of this paper.

\section{Analysis of Perturbed HEE in AdS$_3/$CFT$_2$}

To study the HEE in  AdS$_3/$CFT$_2$, we set $(x^0,x^1)=(t,x)$ in (\ref{adsmet}).
We choose the subsystem A for the entanglement entropy $S_A$ to be an interval
$-l/2+\xi\leq x \leq l/2+\xi$. Then the corresponding minimal surface $\gamma_A$ is
parameterized as \ba t=\mbox{const.},\ \
x=\xi+\f{l}{2}\sin\vp,\ \ z=\f{l}{2}\cos\vp. \label{minth} \ea Then the shift of HEE (\ref{dsa})
reads in terms of the perturbation of the metric $h_{\mu\nu}$: \ba && \Delta
S_A=\f{1}{8G_N}\int^{\pi/2}_{-\pi/2}d\vp
\f{Rl^2\cos\vp}{4z^2}\left[\cos^2\vp h_{xx}+\sin^2\vp
h_{zz}-2\sin\vp\cos\vp h_{xz}\right]\no
&& =\f{R}{8G_N}\int^{\pi/2}_{-\pi/2}d\vp
\cos\vp h_{xx}, \label{pertm} \ea
where we employed the FG gauge in the final expression.

\subsection{HEE in AdS$_3$ Pure Gravity}

As the simplest example in  AdS$_3/$CFT$_2$, we would like to calculate $\Delta S_A$ in the pure Einstein gravity for AdS$_3$.
The equations of motion for the metric perturbation $h_{\mu\nu}$ reads
\ba && \de_z(\de_x h_{tx}-\de_t h_{xx})=0, \no &&
\de_z(\de_x h_{tt}-\de_t h_{tx})=0, \no &&
(\de_z-z\de_z^2)h_{xx}=(\de_z-z\de_z^2)h_{tx}=(\de_z-z\de_z^2)h_{tt}=0,\no
&& \de_z(h_{tt}-h_{xx})-z\de_x^2h_{tt}+2z\de_t\de_x h_{tx}-z\de_t^2
h_{xx}=0. \ea
Note that this system is topological in that there are no propagating
degrees of freedom as usual in three dimensional pure gravity.

By requiring that $h_{\mu\nu}(z,x)$ is order $O(z^2)$ so that only normalizable modes are excited, these
equations can be solved as follows: \ba h_{tt}=h_{xx}=z^2
H(t,x), \ \ \de_t h_{tx}=z^2\de_x H(t,x), \ \ \ \de_x
h_{tx}=z^2\de_t H(t,x), \ea where the function $H(t,x)$ satisfies
\be (\de_t^2-\de_x^2)H(t,x)=0. \label{hwave} \ee

Finally, by using this solution, (\ref{pertm}) is rewritten as
follows: \ba && \Delta
S_A(\xi,l,t)=\f{Rl^2}{32G_N}\int^{\pi/2}_{-\pi/2}d\vp \cos^3\vp\cdot
H\left(t,\xi+\f{l}{2}\sin\vp\right) \no
&& \ \ \ \ \ \ \ \ \ \ \ \ \ \ \ \
=\f{\pi l^2}{4}\int^{\pi/2}_{-\pi/2}d\vp \cos^3\vp\cdot
T_{tt}\left(t,\xi+\f{l}{2}\sin\vp\right) , \label{eepu}
\ea
where we employed the relation $H(t,x)=\f{8\pi G_N}{R}T_{tt}(t,x)$ by setting $d=2$ in
(\ref{EMtensor}).

By taking the Fourier transformation
\be
H(k,t)=\int^{\infty}_{-\infty} dx \ e^{-ikx}H(t,x),
\ee
we obtain
\be
\Delta S_A(k,t,l)=\f{R}{2G_N}\cdot \f{2\sin\f{kl}{2}-lk\cos\f{kl}{2}}{k^3l}\cdot
H(k,t).
\ee
In this way, we find that $\Delta S_A$ is related to the metric via this non-local
transformation. Moreover, it is straightforward to see that it satisfies
\be
\left[\de_l^2+\left(\f{k^2}{4}-\f{2}{l^2}\right)\right]\Delta S_A(k,l,t)=0. \label{kle}
\ee

In summary, we find from (\ref{hwave}) and (\ref{kle}) that $\Delta S_A$ satisfies
 the following ``equations of motion for entanglement entropy'':
\ba
&& (\de_t^2-\de_\xi^2)\Delta S_A(\xi,l,t)=0, \label{eewave} \\
&& \left[\de_l^2-\f{1}{4}\de_{t}^2-\f{2}{l^2}\right]\Delta S_A(\xi,l,t)=0. \label{EOMEE}
\ea
The first equation (\ref{eewave}) shows that the quantum entanglement propagates at the speed of light in the $x$ direction. The second one (\ref{EOMEE}) describes an evolution in the width (or radial) direction, which is analogous to the wave equation in an AdS spacetime. We believe that
the presence of two constraint equations for $\Delta S_A$ as in (\ref{eewave})
and (\ref{EOMEE}) is peculiar to the AdS$_3/$CFT$_2$ duality and this is due to the fact that the gravity does not have propagating degrees in three dimension.

\subsection{HEE from Einstein-Scalar Theory}

Since the pure AdS$_3$ does not have any propagating degrees of freedom, it is more interesting to consider a Einstein-matter theory on AdS$_3$. Though we will not write down all components of Einstein equation, we would like to note that the $tt$ component reads
\begin{equation}
\de_z h_{xx}-z \de_z^2 h_{xx}-2 z T^{(G)}_{tt}=0,
\end{equation}
where $T^{(G)}_{ab}$ is the energy momentum tensor in the $d+1$ dimensional gravity. This should be distinguished from the holographic energy tensor $T_{\mu\nu}$ for the $d$ dimensional dual CFT.
By integrating this equation, the shift of holographic entanglement entropy
$\Delta S_A$ (\ref{pertm}) is expressed as follows:

\begin{equation}
\begin{split}
&\Delta S_A (t,\xi ,l)=\frac{1}{8G_N}\int^{\frac{\pi}{2}}_{-\frac{\pi}{2}}d\varphi R\cos{\varphi}h_{xx}\left(\frac{l}{2}\cos{\varphi},t,\xi+\frac{l}{2}\sin{\varphi}\right) \\
&~~~~~~~~~~~~~=\frac{1}{8G_N}\int ^{\frac{\pi}{2}}_{-\frac{\pi}{2}} d \varphi \cos{\varphi}\bigg{[}-2\int^{\frac{l}{2}\cos(\varphi)}_0 dz'' \int^{z''}_0 dz'T^{(G)}_{tt}\left(z',t,\xi+\frac{l}{2}\sin{\varphi}\right)\cdot \frac{z''}{z'} \\
&~~~~~~~~~~~~~~~~~~~~~~~~~~~~~~~~~~~~~~~~~~~~~~~~~~~~~~~+
\frac{l^2}{4}H\left(t,\xi+\frac{l}{2}\sin{\varphi}\right)\cos^2{\varphi}\bigg{]} .\\
\end{split}
\end{equation}

In particular, we consider the Einstein-scalar theory which is defined by a free scalar field $\phi$ (mass $m$), which is minimally coupled to the Einstein gravity. The Einstein equation is given by (\ref{Ein}) with the energy stress tensor is given by
\be
T^{(G)}_{\mu\nu}=\f{1}{2}\de_\mu\phi \de_\nu\phi-\f{1}{4}g_{\mu\nu}\left[(\de \phi)^2+m^2\phi^2\right],
\ee
where we normalize the scalar field appropriately.

We consider a perturbation of the scalar and AdS$_3$ metric in the FG gauge as follows:
\begin{equation}
\begin{split}
&\phi(z,t,x)=\sqrt{\epsilon}\psi(z,t,x), \\
&g_{\mu \nu}(z,t,x) = \eta_{\mu \nu} + \epsilon h_{\mu \nu}(z,t,x),
\end{split}
\end{equation}
where $\ep$ is an infinitesimally small parameter of the perturbation.

The equation of motion of this perturbation is given by
\begin{equation}
z \left(z \left(\psi^{(0,0,2)}(z,t,x)-\psi^{(0,2,0)}(z,t,x)+\psi^{(2,0,0)}(z,t,x)\right)-\psi^{(1,0,0)}(z,t,x)\right)-m^2 R^2 \psi(z,t,x) =0.
\end{equation}
After the Fourier transformation, the normalizable solution for this equation is given by
\begin{equation}
\psi(z,t,x) =z\int^\infty_{-\infty}
d\omega \int^\infty_{-\infty}  dk \f{e^{-i t \omega + i x k}}{\left (\omega^2 -
     k^2 \right)^{\frac {1} {2} \left (m^2 R^2 +
       1 \right)}} \left \langle \mathcal{O}(\omega ,k) \right \rangle J_ {\sqrt {m^2 R^2 + 1}}\left(z \sqrt{\omega ^2-k^2}\right),
\end{equation}
where $\left \langle \mathcal{O}(\omega ,k) \right \rangle$ denotes the expectation value of an operator dual to the scalar field $\phi$ (up to a normalization factor) \cite{GKPW,KW}.

The entanglement entropy is decomposed into two parts:
\begin{equation}
\Delta S_A (\xi,l,t)= \Delta S_A^{P}(\xi,l,t)+ \Delta S_A ^{M}(\xi,l,t).
\end{equation}
The $\Delta S_A^P$ is the same as that in the pure AdS$_3$ case and is given by
(\ref{eepu}). The other term $\Delta S_A^E$ comes from the contribution from the matter field and is expressed as
\begin{equation}
\begin{split}
&\Delta S_A^M(\xi,l,t)=-\frac{R}{4G_N}\int^{\frac{\pi}{2}}_{-\frac{\pi}{2}}d\varphi \int^{\frac{l}{2}\cos{\varphi}}_0dz''\int^{z''}_0dz'T^{(G)}_{tt}
\left(z',\xi+\frac{l}{2}\sin{\varphi},t\right)\cdot
\frac{z''}{z'}\\
&=\frac{R}{16G_N}\int^{\frac{\pi}{2}}_{-\frac{\pi}{2}}d\varphi \cos{\varphi}\int^{\frac{l}{2}\cos{\varphi}}_0dz''\int^{z''}_0dz'\int d\omega_1 d\omega_2 dk_1 dk_2  \\
&~~~~\times  z'' z'\left \langle \mathcal{O}(\omega_1 ,k_1) \right \rangle \left \langle \mathcal{O}(\omega_2,k_2)\right \rangle F(z',\omega_1,\omega_2,k_1,k_2)e^{i\left(-(\omega_1+\omega_2 )t+(k_1+k_2)\left(\xi+\frac{l}{2}\sin{\varphi}\right)\right)} \label{eemt}\\
\end{split}
\end{equation}
$F(z,\omega_1,\omega_2,k_1,k_2)$ is given by
\ba
&& F(z,\omega_1,\omega_2,k_1,k_2)=-\frac {1}{z}\left (\omega_ 1^2 -
     k_ 1^2 \right)^{\frac {-\nu^2} {2}} \left (\omega_ 2^2 -
    k_ 2^2 \right)^{\frac {-\nu^2} {2} } \no
&&  \bigg{[}z \sqrt {\omega_ 1^2 - k_ 1^2} J_ {\nu -  1}\left (z \sqrt {\omega_ 1^2 - k_ 1^2} \right) \bigg{(}\left (1 - \nu \right) J_ {\nu}\left (z \sqrt {\omega_ 2^2 - k_ 2^2} \right) + z \sqrt {\omega_ 2^2 - k_ 2^2} J_ {\nu - 1}\left (z \sqrt {\omega_ 2^2 - k_ 2^2} \right) \bigg{)} \no
&& +J_ {\nu}\left (z\sqrt {\omega_ 1^2 -  k_ 1^2} \right) \bigg{ (}J_ {\nu}\left (z\sqrt {\omega_ 2^2 - k_ 2^2} \right)
\cdot (-k_ 1 k_ 2 z^2 + 2 \nu^2 - 2\nu - \omega_ 1\omega_ 2 z^2  ) \no
&& \ \ \ \ \ - z\sqrt {\omega_ 2^2 - k_ 2^2}\left (\nu - 1 \right)
\cdot J_ {\nu - 1}\left (z\sqrt {\omega_ 2^2 - k_ 2^2} \right) \bigg{)}\bigg{]}, \\
\ea
where we defined $\nu\equiv \s{m^2R^2+1}$.

We can also act a differential operator to get rid of the $\Delta S_A^{P}$ contribution and obtain the following constraint equations
which are satisfied by $\Delta S_A$:
\begin{equation}
\begin{split}
&(\partial_{t}^2-\partial_{\xi}^2)\Delta S_A(l,t,\xi) \\
&=\int d\omega_1
d\omega_2 dk_1 dk_2  F^{(1)}(k_1,k_2,\omega_1,\omega_2,l)\left \langle \mathcal{O}(\omega_1 ,k_1) \right \rangle \left \langle \mathcal{O}(\omega_2,k_2)\right \rangle  e^{-i(\omega_1+\omega_2 )t+i(k_1+k_2)\xi}. \label{eomeet} \\
\end{split}
\end{equation}
\begin{equation}
\begin{split}
&\left(\partial_l^2-\frac{1}{4}\partial_t^2-\frac{2}{l^2}\right)\Delta S_A(l,t,\xi)\\
&\int d\omega_1
d\omega_2 dk_1 dk_2  F^{(2)}(k_1,k_2,\omega_1,\omega_2,l)\left \langle \mathcal{O}(\omega_1 ,k_1) \right \rangle \left \langle \mathcal{O}(\omega_2,k_2)\right \rangle  e^{-i(\omega_1+\omega_2 )t+i(k_1+k_2)\xi}, \label{eomeett} \\
\end{split}
\end{equation}
where we did not write explicitly $F^{(1)}$ and $F^{(2)}$ as they can be easily obtained from (\ref{eemt}).

In summary, we have obtained a counterpart of the perturbative Einstein equation in terms of entanglement entropy. Therefore we can regard (\ref{eomeet}) and (\ref{eomeett}) as
the perturbative equations of motion for the entanglement entropy. Once we specify
the expectation value of energy density $T_{tt}(t,x)$ and the scalar operator $O(t,x)$ as functions of $t$ and $x$, then the differential equations (\ref{eomeet}) and (\ref{eomeett}) determine the time evolution of
$\Delta S_A(\xi,l,t)$.

 They share a similar structure with that of the Einstein equation because the left hand side comes from the gravitational physics ($\lr$ entanglement) and the right hand side corresponds to the matter fields ($\lr$ operator expectation values). Note also that they are manifestly gauge invariant (as far as we fix the AdS boundary coordinate), as opposed to the Einstein equation where the metric changes under the coordinate transformation.

One may notice that we only used the Einstein equation involving the space-like component $h_{xx}$. In order to take into account the one for the time-like component we need to consider the entanglement entropy for boosted subsystems. We will leave the details of these for a future problem.

\subsection{Proof of First Law-like Relation}

When $l$ is very small, we can neglect the contributions from the matter fields
$\Delta S_A^{M}$. Therefore in this limit, we find
\be
\Delta
S_A(\xi,l,t)\simeq \f{Rl^2}{24G_N}H(t,\xi). \label{xx}
\ee
We can show this relation (\ref{xx}) confirms the time-dependent version of the first law-like relation (\ref{first}) of the entanglement entropy for AdS$_{3}/$CFT$_{2}$. Indeed, since by definition we find $t_{tt}=H(t,x)$ we can easily reproduce (\ref{first}) from
(\ref{xx})\footnote{note that due to our conventions for the $AdS_3/CFT_2$ case we should replace in (\ref{teff}) $l\rightarrow l/2$.}.

\section{Analysis of Perturbed HEE in AdS$_4/$CFT$_3$}

Now we move on to higher dimensional cases and here we especially consider
a AdS$_4/$CFT$_3$ example. Since the pure Einstein gravity is already dynamical in four or higher dimension, we will concentrate on the pure gravity just for simplicity.

\subsection{Einstein Equation}

We consider a metric perturbation of the AdS$_4$ space.
The metric is again given by (\ref{adsmet}) and (\ref{pert}) with $(x^0,x^1,x^2)=(t,x,y)$. We require that $h_{\mu\nu}$ is order $O(z^3)$ in the limit $z\to 0$ so that we can keep only normalizable deformations.

By performing the Fourier transformation
\be
h_{\mu\nu}(t,x,y,z)=\f{1}{(2\pi)^3}\int d\omega dk_x dk_y e^{-i\omega t+ik_x x+ik_y y}\cdot h_{\mu\nu}(z,\omega,k_x,k_y),
\ee
the perturbative Einstein equation leads to
\be
z\de_z^2 h_{\mu\nu}-2\de_z h_{\mu\nu}+(\omega^2-k^2_x-k^2_y)zh_{\mu\nu}=0,\label{ppeomf}
\ee
for all components of perturbations $h_{tt},h_{tx},h_{ty},h_{xx},h_{xy},h_{yy}$.
This is easily solved as
\be \label{sol4}
h_{\mu\nu}(z,\omega,k_x,k_y)=3\s{\f{\pi}{2}}\cdot H_{\mu\nu}(\omega,k_x,k_y)\cdot (\omega^2-k^2)^{-\f{3}{4}}\cdot z^{\f{3}{2}}\cdot J_{3/2}(\s{\omega^2-k^2}z).
\ee
Note that the explicit form of the Bessel function reads
\be
J_{3/2}(x)=\s{\f{2}{\pi x}}\left(-\cos x+\f{\sin x}{x}\right).
\ee
We can show the following behaviour near the AdS boundary $z\to 0$:
\be\label{zzero}
h_{\mu\nu}(z,\omega,k_x,k_y)\simeq z^3\cdot H_{\mu\nu}
(\omega,k_x,k_y).
\ee
Since we are interested in non-singular and normalizable solutions, we restrict to
the range $\omega^2 >k^2$.

We can show that the Einstein equation is equivalent to (\ref{ppeomf}) and (\ref{eomfc})
\ba
&& h_{tt}=\f{1}{k^2_x-\omega^2}\left[-2k_xk_y h_{xy}+(k_x^2-k_y^2)h_{yy}\right],\no
&& h_{tx}=\f{1}{\omega (\omega^2-k_x^2)}\left[-k_y(\omega^2+k_x^2)h_{xy}+k_x(\omega^2-k_y^2)h_{yy}\right],\no
&& h_{ty}=\f{1}{\omega}(-k_xh_{xy}-k_yh_{yy}),\no
&& h_{xx}=\f{1}{k_x^2-\omega^2}\left[-2k_x k_yh_{xy}+(\omega^2-k_y^2)h_{yy}\right].
\label{eomfc}
\ea
In particular, we can find the relation from (\ref{eomfc})
\be
h_{tt}=h_{xx}+h_{yy}, \label{relah}
\ee
which is interpreted as the traceless condition of energy stress tensor in the dual CFT.

\subsection{Calculations of HEE}

We consider the shift of HEE $\Delta S_A$ by choosing
 the subsystem $A$ to be a disk with a radius $l$. In the pure AdS$_4$, the HEE is computed as the area of the minimal surface $\gamma_A$ given by the half of sphere parameterized by
\ba
x=l\sin\theta \cos\phi+X,\ \ \  y=l\sin\theta \sin\phi+Y, \ \ \ z=l\cos\theta,
\ea
with the range $0<\theta<\pi/2$ and $0<\phi<2\pi$.

The shifted amount of the HEE, denoted as $\Delta S_A$ due to the linear perturbation of the metric is found by using (\ref{dsa}):
\ba
&& \Delta S_A(t,X,Y,l)\no
&&=\f{R^2}{8G_N}\!\int^{2\pi}_0\! d\phi\!\int^{\pi/2}_0\! d\theta\!
\f{\sin\theta}{\cos^2\theta} \left[(1-\sin^2\theta \cos^2\phi)h_{xx}-2h_{xy}\sin^2\theta\cos\phi\sin\phi
+(1-\sin^2\theta\sin^2\phi)h_{yy}\right].\label{HEFS}  \no
\ea
We take the Fourier transformation of $\Delta S_A(t,X,Y,l)$ with respect to
$t,X,Y$, which is denoted by $\Delta S_A(\omega,k_x,k_y,l)$:
\ba
&& \Delta S_A(t,X,Y,l)=\f{1}{(2\pi)^3}\int d\omega dk_x dk_y  e^{-i\omega t+ik_x X+ik_y Y},\Delta S_A(\omega,k_x,k_y,l)\no
&& \Delta S_A(\omega,k_x,k_y,l)=\int dt dX dY e^{i\omega t-ik_x X-ik_y Y}\Delta S_A(t,X,Y,l).
\ea

Then we find
\ba
&& \Delta S_A(\omega,k_x,k_y,l) \no
&& =\f{3R^2}{8G_N}\s{\f{\pi}{2}}\int^{\pi/2}_0 d\theta \f{\sin\theta}{\cos^2\theta}
\int^{2\pi}_0 d\phi I(\phi,\theta)\cdot e^{il\sin\theta(k_x\cos\phi+k_y\sin\phi)}\no
&&\ \ \ \ \ \times \left(\f{l\cos\theta}{\s{\omega^2-k^2}}\right)^{3/2}\cdot J_{3/2}(l\cos\theta\s{\omega^2-k^2}),\ \ \  \label{HEEFR}
\ea
where we defined
\be
I(\phi,\theta)=(1-\sin^2\theta \cos^2\phi) H_{xx}(\omega,k)-2\sin^2\theta\cos\phi\sin\phi H_{xy}(\omega,k)
+(1-\sin^2\theta\sin^2\phi) H_{yy}(\omega,k).
\ee

We can perform $\phi$ integral by using the formula of Bessel functions
\be
J_n (z)=\f{1}{2\pi}\int^{2\pi}_0d\phi e^{in\phi-iz\sin\phi},
\ee
as follows:
\ba
&& \f{1}{2\pi}\int^{2\pi}_0 d\phi I(\phi,\theta)\cdot e^{il\sin\theta(k_x\cos\phi+k_y\sin\phi)} \no
&& =(H_{xx}+H_{yy})\left(1-\f{\sin^2\theta}{2}\right)J_0(kl\sin\theta)
-\f{iH_{xy}}{2}\sin^2\theta\left(e^{2i\ap} J_{-2}(kl\sin\theta)-e^{-2i\ap}J_{2}(kl\sin\theta)\right) \no
&&
+\f{\sin^2\theta}{4}(H_{yy}-H_{xx})\left(e^{2i\ap}J_{-2}(kl\sin\theta)
+e^{-2i\ap}J_{2}(kl\sin\theta)\right), \no
&&= \left[\left(1-\f{\sin^2\theta}{2}\right)J_0(kl\sin\theta)
+\f{\sin^2\theta}{k^2}\left(\omega^2-\f{k^2}{2}\right)J_2(kl\sin\theta)\right]H_{tt}.
\ea
Here
the angle $\ap$ was introduced such that $k_x\cos\phi+k_y\sin\phi=-k\sin(\phi+\ap)$ i.e. $\sin\ap=-k_x/k$ and $\cos\ap=-k_y/k$; we employed (\ref{eomfc}) to get the final equation.

Thus we can express $\Delta S_A(\omega,k_x,k_y,l)$ in term of the (holographic) energy
stress tensor $T_{tt}(\omega,k_x,k_y)$ as follows
\ba
&& \Delta S_A(\omega,k_x,k_y,l) \no
&& =\f{2\s{2}\pi^{5/2}l^{3/2}}{(\omega^2-k^2)^{3/4}}\cdot T_{tt}(\omega,k_x,k_y)\cdot \int^{\pi/2}_0 d\theta \f{\sin\theta}{\s{\cos\theta}}Q(\theta), \label{Trel}
\ea
where $Q(l)$ is defined by
\be
Q(\theta)=J_{3/2}(l\cos\theta\s{\omega^2-k^2})\cdot \left[\left(1-\f{\sin^2\theta}{2}\right)J_0(kl\sin\theta)
+\f{\sin^2\theta}{k^2}\left(\omega^2-\f{k^2}{2}\right)J_2(kl\sin\theta)\right].
\ee

\subsection{First law-like Relation and Translationally
Invariant Limit}

By taking the limit $l\to 0$ in (\ref{HEFS}), we obtain
\ba
&& \Delta S_A=\f{\pi R^2 l^3}{4G_N}\int^{\pi/2}_0 d\theta \sin\theta\cos\theta (1-\sin^2\theta/2)(H_{xx}+H_{yy}),\no
&& \ \ \ =\f{3\pi R^2 l^3}{32 G_N}H_{tt},
\ea
where we employed the relation (\ref{relah}). On the other hand the total energy in $A$ reads
\be
\Delta E_A=\pi l^2 T_{tt}=\f{3R^2l^2}{16G_N}H_{tt},
\ee
by using (\ref{EMtensor}).
Therefore we can confirm the first law-like relation (\ref{first}) with $T_{eff}=\f{2}{\pi l}$.
In other words the energy density and $\Delta S_A$ are related to each other via
\be
\Delta S_A=\f{\pi^2l^3}{2}T_{tt}. \label{eep}
\ee

Moreover, it is also intriguing to note that we can obtain the same result (\ref{eep}) when we take the translationally invariant limit $k_x,k_y \to 0$ even if we keep $l$ and
$\omega$ finite. This fact can be shown by explicitly evaluating $\Delta S_A$ as follows:
\ba
&& \Delta S_A(\omega,x,y,l) \no
&&  =\f{3\pi R^2 l^{3/2}}{8G_N\omega^{3/2}}\s{\f{\pi}{2}}\int^{\pi/2}_0 d\theta \f{\sin\theta}{\s{\cos\theta}} \left(2-\sin^2\theta+\f{l^2\omega^2\sin^4\theta}{4}\right) J_{3/2}(l\omega \cos\theta) (H_{xx}+H_{yy}),\no
&& =\f{3\pi R^2 l^3}{32G_N}H_{tt}. \label{kzf}
\ea

\subsection{Metric Shift in CFTs}

In the main part of this paper we have considered only normalizable perturbations which
correspond to the dynamical change of state in a given CFT. However, we would like to
study non-normalizable modes only in this subsection. Indeed, when we consider the time-dependent background corresponding to quantum quenches induced by a sudden change of the metric, we need to take into account the non-normalizable modes\footnote{If we consider quantum quenches induced by excitations of matter fields
such as scalar fields, we will have milder metric backreactions such as the Vaidya
metric. Refer to e.g.\cite{QuenchHol} for recent developments of holographic
quantum quenches.}.

For this, as a solution to the Einstein equation, we assume
\be \label{uno}
h_{\mu\nu}(t,x,y,z)=i\s{\f{\pi}{2}}\cdot \int^\infty_{-\infty} \f{d\omega}{2\pi}
e^{-i\omega t}\ti{H}_{\mu\nu}(\omega)\omega^{3/2}z^{3/2}H^{(1)}_{3/2}(\omega z),
\ee
where the Hankel function is defined by
\be \label{due}
H^{(1)}_{3/2}(z)=\s{\f{2}{\pi z}}\left(\f{-i}{z}-1\right)e^{iz}.
\ee
Near the AdS boundary $z=0$, it approaches
\be
h_{\mu\nu}(t,x,y,z=0)=\int^\infty_{-\infty} \f{d\omega}{2\pi}
e^{-i\omega t}\ti{H}_{\mu\nu}(\omega). \label{frqq}
\ee

Then the HEE behaves like
\ba
&& \Delta S_A(\omega)=\f{i\pi^{3/2} R^2l^{3/2}\omega^{3/2}}{8\s{2}G_N}
\int^{\pi/2-\delta}_0 d\theta \f{\sin\theta}{\s{\cos\theta}}\left(2-\sin^2\theta+\f{l^2\omega^2\sin^4\theta}{4}\right)
H^{(1)}_{3/2}(l\omega \cos\theta)(\ti{H}_{xx}+\ti{H}_{yy}), \no
&& \ \ \ \ =\f{\pi R^2}{32G_N}\left(\f{4+l^2\omega^2}{\delta}+il^3\omega^3\right)
(\ti{H}_{xx}(\omega)+\ti{H}_{yy}(\omega)),
\ea
where $\delta$ is the UV cut off and is related to the lattice spacing $a$ via $\delta=a/l$.

The divergent part of the HEE is simply obtained as
\be
 \Delta S^{div}_A(t)= \f{\pi R^2}{32G_N}(4-l^2\de_t^2)(\ti{H}_{xx}(t)+\ti{H}_{yy}(t))\cdot \f{l}{a},
\ee
where $a$ is the UV cut off (lattice spacing). This agrees with the expectation
from the area law.

The more non-trivial contribution is the finite term, evaluated as follows:
\ba
&&  \Delta S^{finite}_A(t)=\f{\pi R^2l^3}{32G_N}\cdot \de_t^3\left(\ti{H}_{xx}(t,x,y)+\ti{H}_{yy}(t,x,y)\right).
\ea

\section{Analysis of Perturbed HEE in AdS$_5/$CFT$_4$}

Here we analyze the perturbed HEE in the AdS$_5/$CFT$_4$ setup. Since the calculations are parallel with the previous section, our presentation will be brief.

\subsection{Solutions to Perturbative Einstein Equation}

The Einstein equations for the Fourier transformed metric perturbations
\[
h_{\mu\nu}(t,x_1,x_2,x_3,z) = \frac{1}{(2\pi)^4}\int d\omega \; d^3k \; e^{-i\omega t+i k_1 x_1 +i k_2 x_2 +i k_3 x_3 }h_{\mu\nu}(z,\omega,k)
\]

are equivalent to

\be\label{eomfi2}
z \de_z^2 h_{\mu\nu} - 3\de_z h_{\mu\nu} + (\omega^2 -k_1^2-k_2^2-k_3^2)zh_{\mu\nu}=0.
\ee

together with

\begin{eqnarray} \label{c}
h_{tt} &=&\frac{1}{(-k_1^2 + \omega^2)}(2 k_1 k_2 h_{12} + 2 k_1 k_3 h_{13} +
 (- k_1^2+  k_2^2 )h_{22}+ 2 k_2 k_3 h_{23}+( k_3^2 - k_1^2 )h_{33}),\no
h_{t1} &=& \frac {-(k_1^2 k_2+ k_2 \omega^2)h_{12} -
   (k_1^2 k_3 + k_3 \omega^2) h_{13}  +
   (k_1 \omega^2- k_1 k_2^2 ) h_{22} - 2 k_1 k_2 k_3 h_{23} +
  ( k_1 \omega^2 - k_1 k_3^2 )h_{33} }{\omega (- k_1^2 + \omega^2)}, \no
h_{t2} &=& -\frac{1}{\omega}(k_1 h_{12}+ k_2 h_{22}+ k_3 h_{23}), \no
h_{t3} &=& -\frac{1}{\omega}(k_1 h_{13}+ k_2 h_{23}+ k_3 h_{33}), \no
h_{11} &=& \frac{1}{(-k_1^2 + \omega^2)}(2 k_1 k_2 h_{12}+ 2 k_1 k_3 h_{13}+ (k_2^2-
   \omega^2) h_{22} + 2 k_2 k_3 h_{23}+ (k_3^2- \omega^2) h_{33}).
\end{eqnarray}

The normalizable solution of the first equation (\ref{eomfi2}) is
\be \label{sol}
h_{\mu\nu}(z,\omega,k) = 8 (\omega^2 - k^2)^{-1} z^2  J_2(z\sqrt{\omega^2-k^2}) H_{\mu\nu}(\omega,k)= z^4 H_{\mu\nu}(\omega,k) + O(z^6)
\ee

\subsection{Calculation of perturbed HEE}

Consider the HEE $S_A$ for the subsystem $A$ whose boundary is defined by a two dimensional round sphere $S^2$ with radius $l$. The corresponding minimal surface in
the pure AdS$_5$ is parameterized by
\ba
&& z(\theta,\phi, \rho) = l \cos\theta, \ \ \ \  x_1(\theta, \phi, \rho)= l \sin\theta \cos\phi+X_1, \no
&& x_2(\theta, \phi, \rho) = l \sin\theta \sin\phi \cos\rho+X_2, \ \ \ \  x_3(\theta, \phi, \rho)= l \sin\theta \sin\phi \sin\rho+X_3.
\ea

Then the variation of the entanglement entropy is computed as
\ba\label{a}
&&\Delta S_A \no
&&=\frac{R^3}{8G_N}\int_{0}^{\pi}\hspace{-0,2 cm}d\phi \int_{0}^{\pi/2} \hspace{-0,5 cm}d\theta \int^{2\pi}_{0} \hspace{-0,2 cm} d\rho  \frac{\sin^4\theta \sin\phi}{\cos^3\theta}\Biggl[ h_{11}\left( \frac{1}{\sin^2\theta} -  \cos^2\phi\right)  -2\sin\phi\cos (\phi) (h_{12}  \cos\rho  +h_{13} \sin (\rho))\no
&& -2h_{23} \cos\rho \sin\rho \sin^2\phi   +h_{22}\left(\frac{1}{\sin^2\theta}- \cos^2\rho \sin^2\phi  \right) +h_{33}\left(\frac{1}{\sin^2\theta} - \sin^2\phi \sin^2\rho \right) \Biggr].
\ea

After the Fourier transformation
\[
\Delta S(\omega,k,l) = \int dt d^3X \;\Delta S(t,X,l) e^{iwt-ik_i X_i}
\]
we obtain
\begin{equation}\label{dd}
\Delta S(\omega,k,l) =\frac{R^3l^2}{G_N(w^2-k^2)}\int_{0}^{\pi}\hspace{-0,2 cm}d\phi \int_{0}^{\pi/2} \hspace{-0,5 cm}d\theta \int^{2\pi}_{0} \hspace{-0,2 cm} d\rho  \frac{\sin^4\theta \sin\phi}{\cos\theta}  J_2(l \cos\theta\sqrt{\omega^2-k^2})\times
\end{equation}
\[
\times e^{i l\sin\theta(k_1 \cos\phi+k_2 \sin\phi\cos\rho+k_3 \sin\phi\sin\rho )}P(\omega,k,\phi,\theta,\rho)
\]
with
\[
P(\omega,k,\phi,\theta,\rho)=  H_{11}(\omega,k)( \frac{1}{\sin^2\theta} -  \cos^2\phi)  -2\sin\phi\cos \phi (H_{12}(\omega,k)  \cos\rho  +H_{13}(\omega,k) \sin (\rho))-
\]
\[
-2H_{23}(\omega,k) \cos\rho \sin\rho \sin^2\phi   +H_{22}(\omega,k)(\frac{1}{\sin^2\theta}- \cos^2\rho \sin^2\phi  ) +H_{33}(\omega,k)(\frac{1}{\sin^2\theta} - \sin^2\phi \sin^2\rho )
\]

\subsection{First Law-like Relation}

Using (\ref{a}) in the limit of small $l$ (\ref{expa}),
we can perform the integrals as:
\ba\label{b}
&& \Delta S_A =  -\frac{l^4 R^3\pi}{8G_N}\int_{0}^{\pi}\hspace{-0,2 cm} d \phi \int_{0}^{\pi/2} \hspace{-0,5 cm} d \theta  \sin^2\theta \sin\phi\cos\theta [ -2(H_{11}+H_{22}+H_{33})+2H_{11} \sin^2\theta\cos^2\phi  \no
&& \ \ \ \ \ \ +(H_{22}+H_{33}) \sin^2\phi\sin^2\theta] \no
&& = \frac{2 l^4 R^3\pi}{15G_N}(H_{11}+H_{22}+H_{33}).
\ea

Since from equation (\ref{c}) it is immediate to obtain $H_{tt}=H_{11}+H_{22}+H_{33}$,
we can finally show the relation :
\begin{equation}\label{e}
\Delta S_A = \frac{2 l^4 R^3\pi}{15G_N}H_{tt}. 
\end{equation}
This proves the first law-like equation (\ref{first}) by using (\ref{EMtensor}).

More generally, we can show the relation (\ref{e}) if we take $k_{1,2,3}\to 0$ limit with keeping $\omega$ and $l$ finite as in the AdS$_4$ case.

\section{Entanglement Density and Its Perturbation}

Since the entanglement entropy measures the entanglement between a certain region and
its outside, it is a highly non-local quantity. When we are interested in local physics, the entanglement entropy sometimes smears its essential effect. To improve this, the quantity called entanglement density
was introduced in \cite{NNT} for two dimensional
field theories. This quantity measures the entanglement between infinitesimally small two regions and is guaranteed to be positive due to the strong subadditivity relation.
We will generalize this quantity in higher dimensional field theories and study the behavior of entanglement density in excited states by using the perturbative
Einstein equation in the holographic description.

\subsection{Entanglement Density in Two Dimension}

Let us start with the entanglement density in two dimensions introduced in \cite{NNT}.
This is defined by
\be
n(l,\xi,t)=\f{1}{2}\left(\f{1}{4}\de_\xi^2-\de_l^2\right)S_A(l,\xi,t),
\ee
where we chose the subsystem $A$ to be an interval such that its width is $l$ and
its center is at $x=\xi$ as in the holographic description (\ref{minth}).
We also write the shift of entanglement density compared to that of the ground state
as $\Delta n$.

In our holographic setup of the pure gravity on AdS$_3$, this is evaluated as follows
\ba
&& \Delta n(k,t,l)=\f{R}{2G_N}\cdot  \f{-2\sin\f{kl}{2}+lk\cos\f{kl}{2}}{k^3l^3}\cdot
H(k,t)\no
&& \ \ \ \ \ \ \ \ \ \ \ \ \ \ =4\pi \cdot  \f{-2\sin\f{kl}{2}+lk\cos\f{kl}{2}}{k^3l^3}\cdot T_{tt}(k,t). \label{edtw}
\ea
Thus in this case we find the following simple relation is satisfied for the small excitations
around the ground state:
\be
\Delta S_A(\xi,t,l)+l^2\cdot \Delta n(\xi,t,l)=0.
\ee

In the gravity dual of the Einstein scalar theory, we can show from (\ref{eomeett}) that this relation is modified:
\ba
&& \Delta S_A(\xi,t,l)+l^2\cdot \Delta n(\xi,t,l)\no
&&= -\f{l^2}{2}\int d\omega_1
d\omega_2 dk_1 dk_2  \left(F^{(2)}+\f{F^{(1)}}{4}\right)\left \langle \mathcal{O}(\omega_1 ,k_1) \right \rangle \left \langle \mathcal{O}(\omega_2,k_2)\right \rangle  e^{-i(\omega_1+\omega_2 )t+i(k_1+k_2)\xi}.\nonumber
\ea

The first law-like relation (\ref{xx}) can be expressed in terms of the entanglement
density as
\be
\lim_{l\to 0}\Delta n(\xi,t,l)=-\f{\pi}{3}T_{tt}(\xi,t),
\ee
 which can also be seen easily by taking $l\to 0$ limit in (\ref{edtw}). In this way, the entanglement density is equivalent to the energy density when its width $l$ is vanishing. Though this was already noted in \cite{NNT} when $T_{tt}(\xi,t)$ does not depend on $\xi$ and $t$, here we gave a proof of this for more general cases.

\subsection{Entanglement Density in Higher Dimensions}

Now we would like to define the entanglement density for field theories in
higher dimensions. Consider a $d(\geq 3)$ dimensional Minkowski spacetime and define its coordinate $(t,\vec{x})$ ($\vec{x}$ is a $d-1$ dimensional vector). We specify an arbitrarily given subsystem $A$ by $\vec{x}=\vec{x}(\zeta)$, where $\zeta^\ap$
($\ap=1,2,\ddd, d-2$) is the coordinate of the boundary $\de A$ of the subsystem $A$.
Remember that $A$ is defined on a certain time slice $t=$const.
The unit normal vector at a point on $\de A$ (toward the outside direction)
is denoted by $\vec{N}(\zeta)$.

We deform the subsystem $A$ as
\be
\vec{x}=\vec{x}(\zeta)+\delta x_n(\zeta)\cdot \vec{N}(\zeta). \label{deformA}
\ee
In this setup we define the entanglement density by
\be
n_A(\zeta,\zeta')=-\f{\delta^2 S_A}{\delta x_n(\zeta)\delta x_n(\zeta')},
\ee
assuming $\zeta\neq \zeta'$. Notice that this quantity depends not only on
$\zeta$ and $\zeta'$ but also on the choice of subsystem $A$ itself. In this sense,
the entanglement density in higher dimensions ($d>2$) is not completely a local
quantity.

We can show that this quantity is non-negative due to the strong subadditivity of the
entanglement entropy \cite{SSA,HSSA}. To see this, let us assume $\delta x_n=\delta x^{(1)}_n+\delta x^{(2)}_n$, such that only one of $\delta x^{(1)}_n$ and $\delta x^{(2)}_n$ can be
non-zero (or equally $\delta x^{(1)}_n\cdot \delta x^{(2)}_n=0$) for any points on $\de A$. Then the strong subadditivity leads to
\be
S(\vec{x}+\delta x^{(1)}_n)+S(\vec{x}+\delta x^{(2)}_n)\geq
S(\vec{x})+S(\vec{x}+\delta x^{(1)}_n+\delta x^{(2)}_n), \label{ssac}
\ee
where $S(\vec{x}+\delta x^{(1)}_n)$ denotes $S_A$ for the subsystem $A$ defined by $\vec{x}=\vec{x}(\zeta)+\delta x^{(1)}_n(\zeta)\cdot \vec{N}(\zeta)$. By performing
the Taylor expansion of (\ref{ssac}), we find that $n_A(\zeta,\zeta')$ is positive.

\subsection{Holographic Calculation in Pure AdS}

Now we present a holographic calculation of entanglement density in any dimension.
We take the subsystem $A$ to be a half space defined by $y(\equiv x_{d-1})<0$. We perform an
infinitesimal deformation of $A$ (\ref{deformA}) by choosing $\zeta^\ap=x^\ap$ ($\ap=1,2,\ddd,d-2$). In this case, the normal vector $\vec{N}(\zeta)$ is in the
$y$ direction and thus the boundary $\de A$ of the subsystem $A$ is simply expressed as
\be
y=\delta x_n(x^\ap). \label{deformB}
\ee

We would like to calculate $S_A$ for this deformed subsystem by using the AdS/CFT.
Consider the pure AdS$_{d+1}$ in the Poincare coordinate $(z,t,x_1,…,x_{d-2},y)$, whose metric is written as
\begin{equation}
ds^2=R^2\frac{dz^2 - dt^2 +  \sum _{\ap= 1}^{d-2} dx_\ap ^2 + dy^2}{z^2}. \label{POads}
\end{equation}

Before we do the deformation (\ref{deformB}), the minimal surface responsible for $S_A$ is simply given by the plane $y=0$ in (\ref{POads}). Therefore, we can describe the minimal surface after the  deformation (\ref{deformB}) as
\be
 y = f(z,\vec{x}), \label{msf}
\ee
where $\vec{x}=(x_1,x_2,\ddd,x_{d-2})$.

To calculate the HEE , we take a variation of the area functional in order to find the minimal surface:
\begin{equation}
S = \frac{R^{d-1}}{4 G_N} \int \frac{dx^{d-2}dz}{z^{d-1}} \sqrt{1 +  \left(\frac{\partial f}{\partial z} \right)^2  + \sum _{i = 1}^{d-2}  \left(\frac{\partial f}{\partial x_i} \right)^2   }.
\end{equation}

Up to the second order expansion of $f$,  $S_A$ is approximated as
\begin{equation}
S = \frac{R^{d-1}}{4 G_N} \int \frac{dx^{d-2}dz}{z^{d-1}} \left(1 +\frac{1}{2}  \left(\frac{\partial f}{\partial z} \right) ^2 + \frac{1}{2} \sum _{i = 1}^{d-2}  \left(\frac{\partial f}{\partial x_i} \right)^2   \right). \label{areaap}
\end{equation}

By taking a variation with respect to $f$, we find the following equation

\begin{equation}
z^{d-1}\frac{\partial f}{\partial z} \left( \frac{1}{z^{d-1}} \frac{\partial f}{\partial z} \right) + \sum _{i = 1}^{d-2} \frac{\partial ^2 f}{\partial x_i ^2} = 0.
\end{equation}

By performing the Fourier transformation of $ f(z, \vec{x}) $, we obtain

\begin{equation}
\frac{d^2 f}{d z^2} - \frac{d-1}{z}\frac{d f}{d z} - k^2 f= 0, \ \ \  ( k = | \vec{k}| ).
\end{equation}

We can find solutions to this equation that do not diverge as we take $z$ to infinity:
\begin{equation}
f(z) = z^{\frac{d}{2}} K_{\frac{d}{2}} (kz).
\end{equation}

Also we can find the solution that satisfies  $f(0, \vec{x})  = \xi(x)$ at the AdS
boundary is given by

\begin{equation}
 f(z, \vec{x} ) = \frac{\Gamma (d-1 )}{\pi ^{\frac{d-2}{2}} \Gamma (\frac{d}{2})} \int d \tilde{x}^{d-2} \frac{\xi (\vec{\tilde{x}}) z^{d}}{((\vec{x} - \vec{\tilde{x}})^2 + z^2 )^{d-1}}. \label{minfd}
\end{equation}

Now let us substitute this to the area functional (\ref{areaap}). We ignore the divergent part because its dependence of $\xi(x)$ is local and we are not interested in it.
Therefore we concentrate on the finite part given by
\begin{equation}
S_A^{finite} = \frac{R^{d-1}}{4 G_N} \int \frac{d \vec{x}dz}{z^{d-1}} \left( \frac{1}{2} \left( \frac{\partial f}{\partial z} \right)^2 + \frac{1}{2} \sum _{i = 1}^{d-2} \left(\frac{\partial f}{\partial x_i} \right)^2  \right).
\end{equation}

This is evaluated as follows:
\begin{equation}
 S_A^{finite} = - \frac{ R^{d-1} d \Gamma (d-1 )}{8 \pi ^{\frac{d-2}{2}} \Gamma (\frac{d}{2}) G_N} \int dx_1^{d-2} dx_2^{d-2} \frac{\xi (\vec{x_1}) \xi (\vec{x_2})}{(\vec{x_1} - \vec{x_2})^{2(d-1)}}.
\end{equation}

In this way, we obtain the entanglement density from the AdS/CFT:
\begin{align}
n(\vec{x_1}, \vec{x_2})_A &= - \frac{\delta ^2 S_A}{\delta \xi(\vec{x_1}) \delta \xi (\vec{x_2})}  \nonumber  \\
&= \frac{R^{d-1} d\cdot \Gamma (d-1 )}{4 \pi ^{\frac{d-2}{2}} \Gamma (\frac{d}{2}) G_N} \frac{1}{(\vec{x_1} - \vec{x_2})^{2(d-1)}}. \label{hed}
\end{align}

When $d=3$ the above computation has already been done in \cite{SeYo} in the context of
holographic Wilson loops. Moreover, notice that our calculation is essentially the same as that of a two point function of a marginal scalar operator (dual to a massless scalar field) in the AdS$_{d}/$CFT$_{d-1}$ setup \cite{GKPW}.
This suggests that the result (\ref{hed}) may be interpreted as a two point function of certain operators with the conformal dimension $d-1$ in the dual CFT$_{d}$. We naturally expect that such an operator is related to the energy stress tensor. It will be an intriguing future problem to work out this precisely.

It is also possible to calculate the entanglement density when we choose the subsystem $A$ to be a small perturbation of a round ball $B^{d-1}$. The corresponding minimal surfaces can be obtained by performing the conformal transformation in the AdS space (see \cite{CMAP}):
\ba
&& x'^\mu=\f{x^\mu+c^\mu(x^2+z^2)}{1+2c\cdot x+c^2(x^2+z^2)}, \no
&& z'=\f{z}{1+2c\cdot x+c^2(x^2+z^2)}.
\ea

The result is simply given by
\be
n(\vec{\Omega_1},\vec{\Omega_2})=-\f{\delta^2 S_A}{\delta r(\vec{\Omega}_1)\delta r(\vec{\Omega}_2)}=\frac{R^{d-1} d\cdot \Gamma (d-1 )}{4l^2 \pi ^{\frac{d-2}{2}} \Gamma (\frac{d}{2}) G_N} \cdot \f{1}{\left|\vec{\Omega}_1-\vec{\Omega}_2\right|^{2(d-1)}},
\ee
where the unit vectors $\Omega_{1,2}$ parameterize the two positions on S$^{d-2}$ and $r$ is the radial coordinate orthogonal to S$^{d-2}$.

\subsection{Entanglement Density and Metric Perturbation in Higher Dimensions}

The main motivation to study the entanglement density in this paper is to
understand its dynamical properties by using the holographic calculations.
Here we would like to study this in the setup of AdS$_4/$CFT$_3$. In particular, we will holographically compute the change of entanglement density in the presence of small perturbation of the metric compared with the pure AdS result (\ref{hed}) by looking at the pure gravity in four dimensions. We again choose $A$ to be a small perturbation around the
half space.

In the metric background (\ref{adsmet}) with the perturbation (\ref{pert}) at $d=3$,
the shift of HEE $\Delta S_A$ for the surface (\ref{msf}) is given by the following expression up to the linear order of $h_{\mu\nu}$ and the quadratic order of $f$:
\be
\Delta S_A=\f{R^2}{4G_N}\int\f{dxdz}{z^2}\left[\f{h_{xx}}{2}+h_{xy}\f{\de f}{\de x}
+\f{-h_{xx}+2h_{yy}}{4}\left(\f{\de f}{\de x}\right)^2
+\f{h_{xx}+2h_{yy}}{4}\left(\f{\de f}{\de z}\right)^2\right]. \label{HEE}
\ee
The minimal surface $f$ can be found as (\ref{minfd}) by setting $d=3$.

We can obviously divide (\ref{HEE}) into three parts $\Delta S_A=\Delta S_A^{(0)}+\Delta S_A^{(1)}+\Delta S_A^{(2)}$, where $\Delta S_A^{(i)}$ is defined to be the term which involves $i$-th power of $f$. We can find from $\Delta S_A^{(1)}$
\be
\f{\delta \Delta S_A}{\delta \xi(x_1)}=\f{R^2}{2G_N\pi}\int\f{dxdz}{z^2}
h_{xy}\cdot \left(\f{4z^3(x_1-x)}{((x_1-x)^2+z^2)^3}
\right).
\ee
In the above expression, the metric perturbations $h_{\mu\nu}$ are evaluated at
any constant values of $y$ and $t$, corresponding to
 the definition of the subsystem A.

Moreover, we can evaluate the shift of the entanglement density from $\Delta S_A^{(2)}$ as follows:
\ba
&& \Delta n_A(t,y,x_1,x_2) \no
&& =-\f{\delta^2 \Delta S_A}{\delta \xi(x_1)\delta \xi(x_2)}\no
&& = -\f{R^2}{2\pi^2 G_N}\int^{\infty}_0 dz\int^{\infty}_{-\infty}
dx \left[A(x,z) h_{yy}(t,y,z)+B(x,z)(h_{xx}(t,y,z)+h_{yy}(t,y,z))\right], \label{dnab}
\ea
where $A$ and $B$ are given by
\ba
&& A(x,z)
=z^2\f{\left(3(x-x_1)^2-z^2\right)\left(3(x-x_2)^2-z^2\right)+48z^2(x-x_1)(x-x_2)}
{\left((x-x_1)^2+z^2\right)^3\left((x-x_2)^2+z^2\right)^3},\no
&& B(x,z)= z^2\f{\left(3(x-x_1)^2-z^2\right)\left(3(x-x_2)^2-z^2\right)-16z^2(x-x_1)(x-x_2)}
{\left((x-x_1)^2+z^2\right)^3\left((x-x_2)^2+z^2\right)^3}.
\ea

By using the expression (\ref{dnab}) we would like to study the relation between the
entanglement density $\Delta n_A$ and energy momentum tensor
$T_{\mu\nu}$. The latter is related to the asymptotic
behavior of the metric via (\ref{expa}) and (\ref{EMtensor}). The full metric perturbation is found as (\ref{sol4}) by solving the Einstein equation.

In order to proceed analytically we assume a time-dependent but translationally invariant perturbation of the pure AdS$_4$, which is explicitly expressed as (\ref{prtft}).
By taking the Fourier transformation of $\Delta n_A$ with respect to the time $t$, we finally obtain the following result:

\ba
&& \Delta n_A(\omega,y,x_1,x_2) \no
&& =-4\frac{\s{2}}{\sqrt{\pi}}\cdot |\omega|^{-3/2}\cdot
\int^{\infty}_0 dz\int^{\infty}_{-\infty}
dx \left[A(x,z) T_{yy}(\omega)+B(x,z)(T_{xx}(\omega)+T_{yy}(\omega))\right]
z^{3/2}J_{3/2}(|\omega|z),\no
&& =\f{\pi}{16}e^{-l|\omega|/2}|\omega|\cdot \left((l|\omega|-12)T_{yy}(\omega)
+2T_{tt}(\omega)\right),
\ea
where we set $l=|x_1-x_2|$.

We can Fourier transform $\Delta n_A(\omega,y,x_1,x_2)$ and express the result as a perturbation at time $\tilde{t}$ carried by a Green function to time $t$, obtaining

\be
 \Delta n_A(t,y,x_1,x_2)  =\int d\ti{t}\ (G_{tt}(t-\ti{t})\cdot T_{tt}(\ti{t})+G_{yy}(t-\ti{t})\cdot T_{yy}(\ti{t}) )
\ee

 with

\be
G_{yy}(T) = -2\cdot\frac{  l^4 + 6 l^2 T^2 - 24 T^4}{(l^2 + 4 T^2)^3},\ \ \ \
G_{tt}(T) = \frac{ (l^2-4 T^2) }{2(l^2 + 4 T^2)^2}
\ee

\section{Conclusions}

In this paper, we studied the dynamics of holographic entanglement entropy (HEE) induced by a metric perturbation around the pure AdS spacetime. Its dynamics is constrained by the perturbative Einstein equation. First we solved this to express the HEE in terms of the energy stress tensor in the dual CFT. Next we obtained constraint differential equations (\ref{eomeet}) and (\ref{eomeett}), which are satisfied by
the perturbation of HEE for the Einstein-Scalar theory in AdS$_3$. We can regard this as a counterpart of Einstein equation from the viewpoint of the CFT. 

Moreover, we evaluated the evolution of entanglement entropy in the pure gravity theory on AdS$_4$ and AdS$_5$. In particular we found that the increased amount of HEE $\Delta S_A$  is given in terms of the energy density $T_{tt}$ via a non-local transformation as in the equation (\ref{Trel}). 

It will be an important and interesting future problem to find a universal constraint equation for general setups in higher dimensions and to work out how to take into account higher order perturbations systemically in these arguments.

From the CFT viewpoint, our analysis predicts the behavior of entanglement entropy for excited states. As we explicitly show for AdS$_3$ and AdS$_4$ setups, the entanglement propagates at the speed of light and this is dual to the propagation of gravitational waves in the bulk AdS.

At the same time our analysis gives a further support of the first law-like relation
\cite{BNTU} when we choose the subsystem to be a round ball. Our result shows that we get the first law-like relation if the spatially inhomogeneous modulation
is small enough compared with the subsystem size i.e. $kl<<1$. 
 In this way we expect a certain fundamental mechanism suggested by this robust first law-like relation. It will be intriguing to find a field theoretic explanation of this property, especially in higher dimensions.

We also present a higher dimensional extension of entanglement density, which was first introduced in \cite{NNT} for two dimensional field theories. We showed that this quantity is non-negative owing to the strong subadditivity. We presented its holographic calculation and extract its perturbation for weakly excited states.

{\it Note Added:} After the present paper was listed on the arXiv, independent papers \cite{Hes,Al,CMy} appeared. In these papers, first law-like relations 
when the subsystem is a strip have also been analyzed. They showed that the variation of entanglement entropy includes other components of energy stress tensors other than 
$T_{tt}$.

\section*{Acknowledgements}

We would like to thank Hong Liu, Robert Myers and Masaki Shigemori for useful comments. We are also very grateful to Jyotirmoy Bhattacharya for an important comment in the first version of this paper. Also we thank Mohsen Alishahiha and Song He for correspondence.
This work of AP is supported by the Japanese Society for the Promotion of Science (JSPS).
TT is supported by JSPS Grant-in-Aid for Challenging
Exploratory Research No.24654057 and JSPS Grant-in-Aid for Scientific
Research (B) No.25287058. TT is also
supported by World Premier International
Research Center Initiative (WPI Initiative) from the Japan Ministry
of Education, Culture, Sports, Science and Technology (MEXT).


\end{document}